\documentclass[%
reprint,
 amsmath,amssymb,
 aps,
]{revtex4-1}
\usepackage[pdftex]{graphicx} 
\usepackage{dcolumn}
\usepackage{bm}

\newcommand{\ave}[1]{\ensuremath{\langle#1\rangle} }
\usepackage[T1]{fontenc}
\usepackage{textcomp}

\begin{document}

\preprint{APS/123-QED}

\title{Intervention Threshold for Epidemic Control in Susceptible-Infected-Recovered Metapopulation Models}

\author{Akari Matsuki$^1$}
\author{Gouhei Tanaka$^{1,2}$}%
\affiliation{%
 $^1$Department of Mathematical Informatics, Graduate School of Information Science and Technology, \\ The University of Tokyo, Tokyo 113-8656, Japan \\
 $^2$Institute for Innovation in International Engineering Education, Graduate School of Engineering, The University of Tokyo, Tokyo 113-8656, Japan 
}%




\date{\today}

\begin{abstract}
Metapopulation epidemic models describe epidemic dynamics in networks of spatially distant patches connected via pathways for migration of individuals. In the present study, we deal with a susceptible-infected-recovered (SIR) metapopulation model where the epidemic process in each patch is represented by an SIR model and the mobility of individuals is assumed to be a homogeneous diffusion. We consider two types of patches including high-risk and low-risk ones under the assumption that a local patch is changed from a high-risk one to a low-risk one by an intervention. We theoretically analyze the intervention threshold which indicates the critical fraction of low-risk patches for preventing a global epidemic outbreak. We show that an intervention targeted to high-degree patches is more effective for epidemic control than a random intervention. The theoretical results are validated by Monte Carlo simulations for synthetic and realistic scale-free patch networks. The theoretical results also reveal that the intervention threshold depends on the human mobility network and the mobility rate. Our approach is useful for exploring better local interventions aimed at containment of epidemics.

\end{abstract}

\maketitle


\section{Introduction}
In the modern age of expanding globalization, epidemic spreading is a serious matter of global public health. Countermeasures, such as vaccinations, antiviral medications, and social distancing, have been practiced for controlling past infectious diseases. However, emerging and re-emerging infectious diseases pose perpetual challenges for controlling them due to environmental changes and diversification of human behavior \cite{fauci2012perpetual,morens2013emerging}. Therefore, it is significant to continuously explore systematic methods for planning effective epidemic control strategies. Mathematical models are powerful tools for understanding epidemic spreading processes which are complex phenomena involved in the type of disease, host immunity, environmental conditions, and human mobility patterns \cite{keeling2011modeling}. Mathematical methods have been widely used to estimate epidemic outcomes and evaluate the effectiveness of preventive measures.  

There are a variety of mathematical models for epidemic spreading, from simple to complex ones. Compartment epidemic models assuming homogeneous mixing of individuals are classical and simple \cite{diekmann2000mathematical}. These models have been extended to more complex and realistic ones by incorporating additional factors, such as social structures \cite{strang1991adding}, spatial structures \cite{riley2015five}, seasonal forcing \cite{tanaka2013effects,bjornstad2016timing,buonomo2018seasonality}, human mobility patterns \cite{barrat2008dynamical,balcan2010modeling,belik2011natural,bajardi2011human,tizzoni2014use,urabe2016parameter}. Metapopulation epidemic models are a class of models that describe epidemic spreading processes in a group of spatially separated patches connected via migration pathways. This framework has been intensively studied to develop analytical methods for global epidemic thresholds \cite{sattenspiel1995structured,keeling2004metapopulation,colizza2007invasion,colizza2008epidemic,belik2011recurrent,balcan2011phase,wang2012safety,apolloni2014metapopulation,wang2014epidemic,tizzoni2015scaling,wang2017interplay,gomez2018critical,granell2018epidemic,soriano2018spreading} and widely applied to understanding actual epidemic outbreaks \cite{cador2016maternally,de2018impact,lee2018metapopulation,meakin2019metapopulation,laager2019metapopulation}. In a metapopulation model, infection and recovery events occur in each patch and migration of individuals potentially causes global epidemic spreading. Metapopulation models have often been employed to consider inhomogeneous mixing of individuals. The SIR metapopulation model with fully connected patches was analyzed to examine the properties of the global basic reproduction number (which is differentiated from the local basic reproduction number in an isolated patch) governing the global epidemic threshold \cite{arino2003multi,allen2008basic}. The global reproduction number can be numerically estimated using the next-generation matrix approach, but a theoretical issue is to derive its explicit expression as a function of system parameters including local epidemiological parameters and human mobility patterns \cite{cross2005duelling}.

Colizza and Vespignani~\cite{colizza2007invasion,colizza2008epidemic} derived an analytical expression of the global invasion threshold (i.e. the global reproduction number) for an SIR metapopulation model with complex patch connectivity under several assumptions. They clarified the effect of heterogeneous network connectivity on the global epidemic threshold. The analysis was conducted under the assumption that the local reproduction number is the same for all the patches. However, the conditions of patches are thought to be heterogeneous in reality. In fact, it was reported that the local reproduction numbers estimated from real data for seasonal influenza are different between local areas \cite{chowell2010reproduction}. The heterogeneity of local reproduction numbers can be partially attributed to the difference in the immunization coverage rates in local areas. Under the patch heterogeneity, the effectiveness of strategic interventions for epidemic control has been evaluated using a susceptible-infected-susceptible (SIS) metapopulation model in our previous study \cite{tanaka2014random}. The result shows that targeted interventions for high-degree patches are more effective than random interventions. However, it is still unclear whether this result holds for other types of metapopulation models. For instance, the SIR epidemic process representing an epidemic outbreak as a transient state is qualitatively different from the SIS epidemic process representing an endemic state as a stationary state. In fact, theoretical approaches are completely different between them; the global epidemic threshold of an SIR metapopulation model is analyzed based on a branching process method \cite{colizza2007invasion,colizza2008epidemic}, while that of an SIS metapopulation model is analyzed based on local stability of the disease-free equilibrium state \cite{colizza2007reaction,tanaka2014random}. Thus, the impact of local intervention in SIR metapopulation models with patch heterogeneity still remains to be studied.

\begin{figure}[t]
\begin{center}
\includegraphics[width=\hsize]{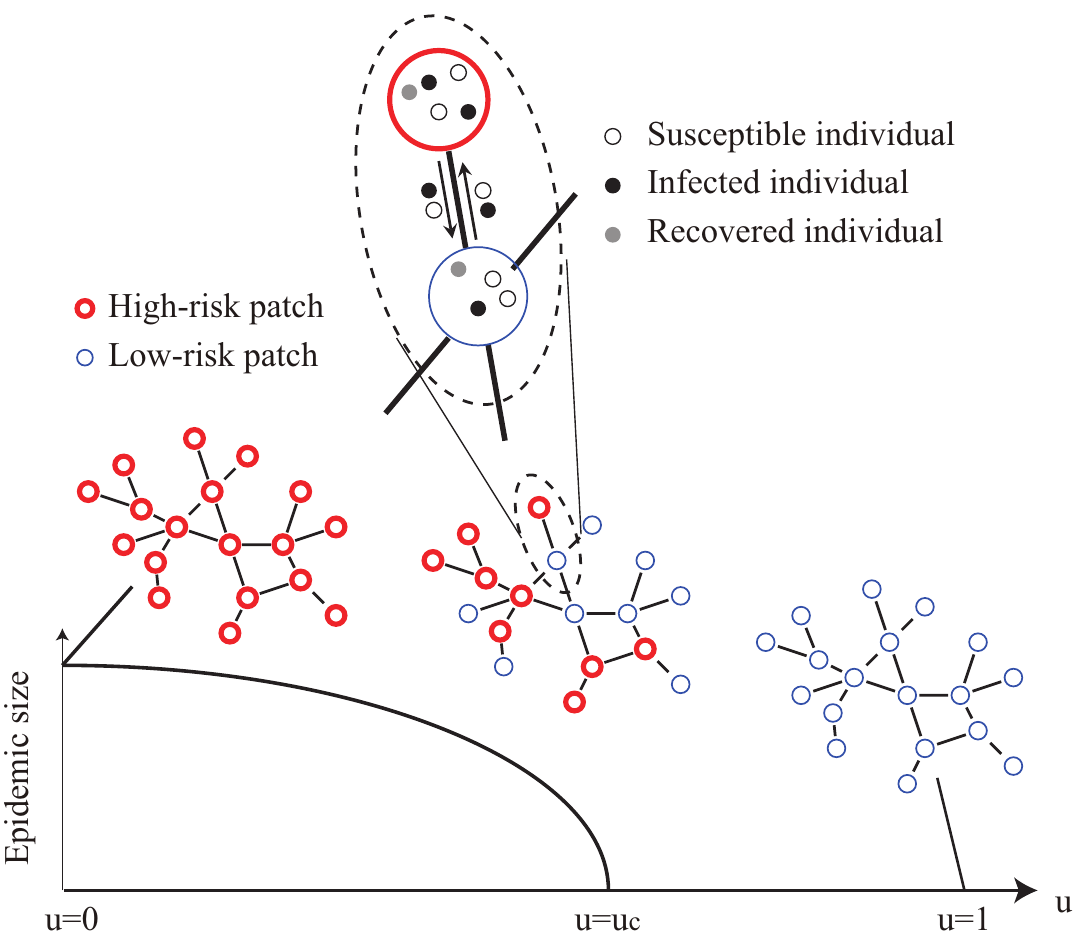}
\caption{Schematic illustration describing the SIR metapopulation model with high-risk patches (red thick circles) and low-risk ones (blue thin circles).}
\label{fig:hlnet}
\end{center}
\end{figure}

In the present study, we aim to analyze the intervention threshold in SIR metapopulation models consisting of high-risk and low-risk patches as shown in Fig.~\ref{fig:hlnet}. This model framework is similar to that considered in Ref.~\cite{tanaka2014random}, but the epidemic process is qualitatively different. We introduce an intervention rate $u$ representing the fraction of low-risk patches that have received intervention. When $u=0$, all the patches are high-risk and a global epidemic outbreak inevitably occurs. In the other extreme case with $u=1$, all the patches are low-risk and a global epidemic outbreak is prevented. Therefore, we can expect that there is a certain critical value $u=u_c \in (0,1)$ (called an {\it intervention threshold}), separating the outbreak and non-outbreak regimes. We use this threshold as a measure to compare different intervention strategies. The smaller the intervention threshold is, the more effective the intervention strategy is. 
The main novelty of this study is to theoretically derive the intervention threshold $u_c$ for random and targeted interventions. Furthermore, our theoretical results are validated by numerical simulations. The comparison of the intervention thresholds shows that targeted intervention for high-degree patches is more effective than random intervention. Our result indicating the effectiveness of targeted intervention in metapopulations reminds us of the effectiveness of targeted immunization in complex contact networks of individuals \cite{pastor2002immunization,pastor2015epidemic}. However, these model frameworks are largely different because dynamical processes within network nodes are considered in metapopulation models but not in contact network models.

In Sec.~\ref{sec:method}, we first introduce the analysis framework proposed in the previous study \cite{colizza2007invasion,colizza2008epidemic} and then describe our approach. In Sec.~\ref{sec:result}, we show theoretical and numerical results. In Sec.~\ref{sec:conclusion}, we conclude this study.



\section{Methods \label{sec:method}}
\subsection{SIR metapopulation model with identical patches \label{subsec:identical}}
We first introduce a method for analyzing a global epidemic threshold in an SIR metapopuation model with identical patches, following Refs.~\cite{colizza2007invasion,colizza2008epidemic}. We will extend this method to the case with non-identical patches in the subsequent section. 

An SIR metapopulation model describes epidemic spreading in a network of spatially separated patches, interconnected with migration pathways. The number of patches is denoted by $V$. The patches are assumed to totally contain a sufficiently large number $N$ of individuals, who are susceptible (S), infected (I), or recovered (R). Epidemic dynamics in each patch follows an SIR process  \cite{kermack1932contributions}: a susceptible individual changes to an infected one with transmission rate $\beta$ when contacting with an infected individual ($\mbox{S}+\mbox{I} \rightarrow 2\mbox{I}$); an infected individual changes to a recovered one with recovery rate $\mu$ ($\mbox{I} \rightarrow \mbox{R}$). We assume homogeneous mixing of individuals in each patch where the local reproduction number is given by $R_0=\beta/\mu$. To allow global epidemic spreading, the local reproduction number $R_0$ needs to be larger than unity. Individuals can migrate from one patch to a neighboring one through the pathway. This is regarded as a diffusion process \cite{colizza2007reaction} and the diffusion rate from a patch is denoted by $p$. Under a homogeneous diffusion process, the diffusion rate from a patch with degree $k$ to one of the neighboring patches (with any degree $k'$) is given by 
\begin{eqnarray}
d_{kk'} &=& \frac{p}{k}. \label{eq:d}
\end{eqnarray}
In a stationary state, the number of individuals in a patch with degree $k$ is obtained as follows \cite{colizza2007invasion,colizza2008epidemic}: 
\begin{eqnarray}
N_k &=& \frac{k}{\langle k \rangle}\bar{N}, \label{eq:N_k}
\end{eqnarray}
where $\bar{N}=N/V$ is the average population size per patch and $\langle k \rangle$ is the mean degree.

We consider an initial condition that an infected individual invades a metapopulation system of susceptible individuals. The total number of infected individuals in a patch is proportional to the stationary population in the patch, described as $\alpha N_{k}$ for a patch with degree $k$, where the coefficient $\alpha$ depends on the type of disease and other factors. The average infection period of an infected individual is given by the inverse of the recovery rate, $\mu^{-1}$. Therefore, if an epidemic occurs in a patch with degree $k$, the average number of infected individuals who move to a neighboring patch with degree $k'$ is represented as follows \cite{colizza2007invasion,colizza2008epidemic}:
\begin{eqnarray}
\lambda_{kk'} &=& d_{kk'}\frac{\alpha N_{k}}{\mu}. \label{eq:lambda}
\end{eqnarray}



%
We focus on the time evolution of the number of ``infected'' patches which are defined as the patches that undergo an outbreak. The analysis is based on the basic branching process \cite{ball1997epidemics,harris2002theory}. We denote by $D_k^0$ the number of infected patches with degree $k$ at generation 0 (i.e. in the beginning of the process). These patches bring about new infected patches with degree $k$ in their neighborhood, the number of which is represented as $D^1_k$ at generation 1. In this way, we define as $D^{n}_{k}$ the number of infected patches with degree $k$ at generation $n$. Assuming that the number of infected patches is sufficiently small in the early stage of the process, we can approximately relate $D^n_k$ to $D^{n-1}_{k}$ as follows \cite{colizza2007invasion,colizza2008epidemic}:
\begin{eqnarray}
&& D_{k}^{n}=\hspace{-1mm} \sum_{k'}D_{k'}^{n-1}(k'-1)P(k|k') \hspace{-1mm} \left(\hspace{-1mm}1-\frac{D_{k}^{n-1}}{V_{k}}\hspace{-1mm}\right) \hspace{-1mm}\left(\hspace{-1mm} 1-R_{0}^{-\lambda_{k'k}} \hspace{-1mm}\right), \label{zennka1} \nonumber \\ \label{eq:D_kn}
\end{eqnarray} 
where $P(k)$ denotes the degree distribution of the patch network and $V_k$ denotes the number of patches with degree $k$. This equation is derived based on the notion that each infected patch with degree $k'$ at generation $(n-1)$ will spread infection in the $(k'-1)$ neighboring patches except the one that originally transmitted infection, the probability that a neighboring patch of a patch with degree $k'$ has degree $k$ is $P(k|k')$, and the probability that the disease does not become extinct when $\lambda_{k'k}$ infected individuals invade in a patch with $R_0$ is given by $(1-R_0^{-\lambda_{k'k}})$ \cite{bailey1975mathematical,Murray}.


For analytical tractability, we deal with special cases under the assumptions of homogeneous diffusion in mobility, uncorrelated patch networks (i.e. without degree-degree correlation), and the local reproduction number close to an epidemic threshold. From Eqs.~(\ref{eq:d})-(\ref{eq:lambda}), the number of seeds of infection is given by 
\begin{eqnarray}
\lambda_{k'k} &=& \frac{p\alpha N_{k'}}{\mu k'} = \frac{p \alpha \bar{N}}{\mu \ave{k}}. \label{eq:lambda2}
\end{eqnarray}
In an uncorrelated patch network, the following equation holds \cite{dorogovtsev2002evolution}: 
\begin{eqnarray}
P(k|k') &=& \frac{kP(k)}{\ave{k}}. \label{eq:P}
\end{eqnarray}
When the local reproduction number is close to the epidemic threshold, i.e. $R_0-1 \ll 1$, the outbreak probability is approximated as follows:
\begin{eqnarray}
1-R_0^{-\lambda_{k'k}} &\simeq & \lambda_{k'k}(R_0-1). \label{eq:1-R0}
\end{eqnarray} 

By substituting Eqs.~(\ref{eq:lambda2})-(\ref{eq:1-R0}) into Eq.~(\ref{zennka1}), we obtain the following equation \cite{colizza2007invasion,colizza2008epidemic}:
\begin{eqnarray}
D_{k}^{n} &=& \frac{p \alpha \bar{N}}{\mu \ave{k}} \frac{kP(k)}{\ave{k}} (R_{0} - 1)\sum_{k'}D_{k'}^{n-1}(k'-1).
\end{eqnarray}
By defining $\Theta^{n} := \sum_{k'} D^{n}_{k'}(k'-1)$, the above equation can be rewritten by the following recurrence formula:
\begin{eqnarray}
\Theta^{n} = \frac{p \alpha \bar{N}}{\mu} \frac{\ave{k^{2}}-\ave{k}}{\ave{k}^{2}} (R_{0} - 1)\Theta^{n-1}.
\end{eqnarray}
The condition that $\Theta^{n}$ does not increase with $n$ is given by \cite{colizza2007invasion,colizza2008epidemic}
\begin{eqnarray}
R_{*} &:=& \frac{p \alpha \bar{N}}{\mu} \frac{\ave{k^{2}}-\ave{k}}{\ave{k}^{2}} (R_{0} - 1) <1, \label{eq:R_*}
\end{eqnarray}
where $R_{*}$ represents the global reproduction number. If $R_{0}$ is close to 1, then $\alpha \simeq 2(R_{0} - 1) / (R_{0})^{2}$ according to Ref.~\cite{Murray}. Using this approximation, Eq.~(\ref{eq:R_*}) is simplified as follows:
\begin{eqnarray}
R_{*} &=& \frac{2p \bar{N}(R_{0}-1)^{2}}{\mu (R_{0})^{2}} \frac{\ave{k^{2}}-\ave{k}}{\ave{k}^{2}}.
\end{eqnarray}

\subsection{SIR metapopulation model with high-risk and low-risk patches \label{subsec:two}}
Extending the framework in Sec.~\ref{subsec:identical}, we analyze an SIR metapopulation model consisting of high-risk and low-risk patches for examining epidemic intervention strategies \cite{allen2007asymptotic,tanaka2014random}. We assume that only a fraction of patches can receive intervention and become low-risk due to budgetary constraints. The local reproduction number in such low-risk patches is denoted by $R_0^L$ and that in the remaining high-risk patches is by $R_0^H (> R_0^L)$. 

Let us define $D^{n}_{k,H}$ and $D^{n}_{k, L}$ as the numbers of infected high-risk and low-risk patches with degree $k$ at generation $n$, respectively. The numbers of individuals who experience the disease during an outbreak in the high-risk and low-risk patches are represented as $\alpha_{H}N_{k}$ and $\alpha_{L}N_{k}$, respectively. The numbers of seeds from high-risk and low-risk patches with degree $k$ are denoted by $\lambda_{k'k}^{H}$ and $\lambda_{k'k}^{L}$, respectively. We define $Q(k)$ as the probability that a randomly chosen patch with degree $k$ is a low-risk one. Considering the transmission of infection from high-risk and low-risk patches separately, the recurrence formulae for $D_{k,H}^{n}$ and $D_{k,L}^{n}$ are written as follows (as in Eq.~(\ref{eq:D_kn})):
\begin{eqnarray}
 D_{k,H}^{n} &=& \sum_{k'}D_{k',H}^{n-1}(k'-1)P(k|k')[1- (R_0^H)^{-\lambda_{k'k}^{H}}] \nonumber \\
& & \times (1-Q(k))\left(1-\frac{D_{k,H}^{n-1}}{V_{k,H}}\right) \nonumber \\
&  & +\sum_{k'} D_{k',L}^{n-1}(k'-1)P(k|k') [1- (R_{0}^{H})^{-\lambda_{k'k}^{L}}] \nonumber\\
& & \times (1-Q(k))\left(1-\frac{D_{k,L}^{n-1}}{V_{k,L}}\right), \label{eq:D_H} 
\end{eqnarray}
\begin{eqnarray}
 D_{k,L}^{n} &=& \sum_{k'}D_{k',H}^{n-1}(k'-1)P(k|k') [1- (R_{0}^{L})^{-\lambda_{k'k}^{H}}] \nonumber \\
& & \times Q(k)\left(1-\frac{D_{k,H}^{n-1}}{V_{k,H}}\right) \nonumber \\
& & + \sum_{k'}D_{k',L}^{n-1}(k'-1)P(k|k') [1- (R_{0}^{L})^{-\lambda_{k'k}^{L}}] \nonumber \\
& & \times Q(k)\left(1-\frac{D_{k,L}^{n-1}}{V_{k,L}}\right), \label{eq:D_L}
\end{eqnarray} 
where $V_{k, H}$ and $V_{k,L}$ represent the numbers of high-risk and low-risk patches with degree $k$, respectively.

As in Eq.~(\ref{eq:lambda2}), we obtain
\begin{eqnarray}
\lambda_{k'k}^{H} &=& \frac{ p \alpha_{H} N_{k'}}{\mu k'} = \frac{ p \alpha_H \bar{N} } {\mu \ave{k}}, \\
\lambda_{k'k}^{L} &=& \frac{ p\alpha_{L} N_{k'} }{\mu k' }= \frac{p \alpha_L \bar{N} } {{\mu} \ave{k}}.
\end{eqnarray}
Assuming $R_{0}^{H} \simeq 1$ and $R_{0}^{L} \simeq 1$, we can use the following approximations (as in Eq.~(\ref{eq:1-R0})):
\begin{eqnarray}
1-(R_0^H)^{-\lambda_{k'k}} &\simeq & \lambda_{k'k}(R_0^H-1), \label{eq:1-R0H} \\
1-(R_0^L)^{-\lambda_{k'k}} &\simeq & \lambda_{k'k}(R_0^L-1). \label{eq:1-R0L}
\end{eqnarray}
In the early stage of the propagation, it follows
\begin{eqnarray}
\left(1-\dfrac{D_{k,H}^{n-1}}{V_{k,H}}\right) \simeq 1 \ \ {\rm and} \left(1-\dfrac{D_{k,L}^{n-1}}{V_{k,L}}\right) \simeq 1.
\end{eqnarray}

By defining 
\begin{eqnarray}
\Theta^{n}_{H} &:=& \sum_{k}D_{k,H}^{n}(k-1), \\
\Theta^{n}_{L} &:=& \sum_{k}D_{k,L}^{n}(k-1), \\
\lbrack k^{\alpha}\rbrack &:=& \sum_{k} k^{\alpha} P(k)Q(k), \label{eq:bracket_k}
\end{eqnarray}
we can rewrite Eqs.~(\ref{eq:D_H})-(\ref{eq:D_L}) as follows:
\begin{eqnarray}
\Theta^{n}_{H} &=& \frac{p \bar{N} \alpha_{H}}{\mu} (R_{0}^{H} - 1) \frac{(\ave{k^{2}}- \ave{k})- (\lbrack k^{2}\rbrack - \lbrack k \rbrack)}{\ave{k}^{2}} \Theta^{n-1}_{H} \nonumber \\
&+& \frac{p \bar{N} \alpha_{L}}{\mu} (R_{0}^{H} - 1) \frac{(\ave{k^{2}}- \ave{k})- (\lbrack k^{2}\rbrack - \lbrack k \rbrack)}{\ave{k}^{2}} \Theta^{n-1}_{L}, \nonumber \\
\\
\Theta^{n}_{L} &=& \frac{p \bar{N} \alpha_{H}}{\mu} (R_{0}^{L} - 1) \frac{\lbrack k^{2}\rbrack - \lbrack k \rbrack}{\ave{k}^{2}} \Theta^{n-1}_{L} \nonumber \\
&+& \frac{p \bar{N} \alpha_{L}}{\mu} (R_{0}^{L} - 1) \frac{\lbrack k^{2}\rbrack - \lbrack k \rbrack}{\ave{k}^{2}} \Theta^{n-1}_{L} . 
\end{eqnarray}
These recurrence equations are simply written as follows:
\begin{eqnarray}
\label{2-theta}
&\left(
	\begin{array}{c}
		\Theta_{H}^{n} \\
		\Theta_{L}^{n}
	\end{array}
\right) 
= J
\left(
	\begin{array}{c}
		\Theta_{H}^{n-1} \\
		\Theta_{L}^{n-1}
	\end{array}
\right),
\end{eqnarray}
where
\begin{eqnarray}
&&J := \hspace{-1mm} \left(\hspace{-2mm}
	\begin{array}{cc}
		\frac{p \bar{N} \alpha_{H} (R_{0}^{H} - 1)}{\mu} (\phi_{1} - \phi_{2}) 
		& \frac{p \bar{N} \alpha_{L} (R_{0}^{H} - 1)}{\mu} (\phi_{1} - \phi_{2}) \\
		\frac{p \bar{N} \alpha_{H} (R_{0}^{L} - 1)}{\mu} \phi_{2} & \frac{p \bar{N} \alpha_{L} (R_{0}^{L} - 1)}{\mu} \phi_{2}
	\end{array}
\right), \nonumber \\
\label{eq:J}\\
&&\phi_{1} := (\ave{k^{2}}- \ave{k}) / \ave{k}^{2}, \label{eq:phi1}\\
&&\phi_{2} := ([k^{2}] - [k]) / \ave{k}^{2}. \label{eq:phi2}
\end{eqnarray}
The eigenvalues of $J$ are given by 0 and
\begin{eqnarray}
\frac{p\bar{N}}{\mu} \{\alpha_{H}(R_{0}^{H}-1)(\phi_{1} - \phi_{2}) + \alpha_{L}(R_{0}^{L}-1)\phi_{2}\}.
\end{eqnarray}
The condition that $\Theta^{n}_{H}$ and $\Theta^{n}_{L}$ do not diverge in the limit of $n \to \infty$ is equivalent to the condition that the absolute values of all the eigenvalues of $J$ are smaller than 1. Hence, the condition that a global outbreak does not occur is given by:
\begin{eqnarray}
R_c &:=& \frac{p\bar{N}}{\mu} \{\alpha_{H}(R_{0}^{H}-1)(\phi_{1} - \phi_{2}) + \alpha_{L}(R_{0}^{L}-1)\phi_{2}\} \nonumber \\
  &<& 1, \label{eq:R_mix}
\end{eqnarray}
where $R_c$ represents the global reproduction number in the case that high-risk and low-risk patches coexist. When $R_0$ is close to unity, $\alpha=2(R_0-1)/R_0^2$. Therefore, the global reproduction number is rewritten as follows:
\begin{eqnarray}
R_c &=& \frac{p\bar{N}}{\mu} \{\psi(R_0^H)(\phi_1-\phi_2)+\psi(R_0^L)\phi_2 \},
\label{eq:R_mix2}
\end{eqnarray}
where $\psi(x) := 2(x-1)^2/x^2$. Based on this formula, we can evaluate the intervention threshold for different strategies as described in Sec.~\ref{subsec:theoanal}.

\subsection{Numerical simulation methods}
We describe numerical methods for simulating epidemic propagation processes in SIR metapopulation models, which are used to validate our theoretical results. The state of each individual is susceptible (S), infected (I), or recovered (R). Initially the population in patch $j$ is set at $N_{j} = k_{j} \bar{N} / \ave{k}$ for $j=1,\ldots,V$. The numbers of susceptible, infected, and recovered individuals in patch $j$ are denoted by $S_j$, $I_j$, and $R_j$, respectively. A patch whose degree is close to $\ave{k}$ is chosen to have ten initial infected individuals. The remaining individuals are susceptible. We consider discrete-time dynamical processes and denote the unitary time step by $\tau$. At each time step, the state of each individual in patch $j$ is probabilistically updated. The update process consists of two stages: epidemic and mobility stages. In the epidemic stage, each susceptible individual turns into an infected one with probability $1-(1-\beta_{j} \tau / N_{j})^{I_{j}}$ and each infected individual turns into a recovered one with probability $\mu \tau$. After all individuals have been updated in the epidemic stage, the mobility stage starts. In the mobility stage, each individual moves to one of the neighboring patches with probability $p\tau$. The above procedure is repeated for all the individuals at each time step and continued for finite time steps until infected individuals disappear.

We set $\bar{N}=1000$, $\tau = 0.1$, and $\mu = 1$ for all the patches, $\beta_{j} = 2$ for high-risk patches, and $\beta_{j} = 1.01$ for low-risk patches, unless otherwise noted. To focus on epidemic spreading in heterogeneous patch networks, we employed synthetic scale-free networks with $V = 200$ patches having degree distribution $P(k) \sim k^{-\gamma}$ with $\gamma = 2.1$ generated by the uncorrelated configuration model \cite{catanzaro2005generation} and the real US airport network having a scale-free property, containing $V = 500$ patches \cite{colizza2007reaction}. We performed 50 simulations with different network realizations for each parameter condition.



\section{Results \label{sec:result}}
First, we theoretically analyze the intervention threshold in SIR metapopulation models with heterogeneously connected patches in Sec.~\ref{subsec:theoanal}. We deal with random and targeted interventions \cite{tanaka2014random}. Then, we numerically validate the theoretical results in Sec.~\ref{sebsec:numerical}.

\subsection{Theoretical results \label{subsec:theoanal}}

For theoretical analysis, we approximate the degree $k$ as a continuous variable by representing its expectation value over many realizations of networks \cite{barabasi2016network}. We consider a probability density function $p(k)$ for a continuous degree distribution, instead of the discrete degree distribution $P(k)$. We also define a probability density function $q(k)$ for a continuous intervention probability, instead of the discrete intervention probability $Q(k)$. Accordingly, the summations with respect to $k$ in the previous section are replaced with integrals over $k$. In particular, we redefine the brackets in Eq.~(\ref{eq:bracket_k}) as follows:
\begin{eqnarray}
\lbrack k^{\alpha}\rbrack &:=& \int_k k^{\alpha} p(k)q(k)dk. \label{eq:bracket_k_cont}
\end{eqnarray}

The total intervention rate $u$ represents the fraction of low-risk patches. For a given $u$, we need to appropriately define $q(k)$ such that
\begin{eqnarray}
0 \le q(k) \le 1, \label{eq:q_range} \\
\int_k q(k) p(k) dk &=& u. \label{eq:q_condition}
\end{eqnarray}

\subsubsection{Threshold for random intervention}
First, we deal with {\it random} intervention, where the low-risk patches are chosen at random. Namely, the probability that a patch is low-risk is constant independently of the patch degree. From Eq.~(\ref{eq:q_condition}), we obtain the probability density function $q(k)$ for the random intervention as follows:
\begin{eqnarray}
q^{\rm rn}(k) &=& u. \label{eq:q_random}
\end{eqnarray}
In this case, we have $\lbrack k \rbrack = u\ave{k}$ from Eq.~(\ref{eq:bracket_k_cont}) and $\phi_{2} = u\phi_{1}$ from Eqs.~(\ref{eq:phi1}), (\ref{eq:phi2}), and (\ref{eq:bracket_k_cont}). Using these equations and Eq.~(\ref{eq:R_mix}), the global reproduction number $R_c$ is described as follows: 
\begin{eqnarray}
R_c^{\rm rn} &=& \frac{p\bar{N}}{\mu}\phi_{1} \{ \alpha_{H}(R_{0}^{H} -1)(1-u) + \alpha_{L}(R_{0}^{L}-1)u \}. \nonumber \\
 \label{eq:R_random}
\end{eqnarray}
By solving $R_c^{\rm rn} = 1$ with respect to $u$,
we obtain the critical intervention threshold as follows:
\begin{eqnarray}
u_c^{\rm rn} &=& \frac{ \alpha_{H}(R_{0}^{H}-1) - \mu/(p\bar{N}\phi_{1}) }{ \alpha_{H}(R_{0}^{H} -1) - \alpha_{L}(R_{0}^{L}-1) }, \label{eq:uc_random}
\end{eqnarray}
above which a global epidemic outbreak is prevented.

\begin{figure}[t]
\begin{center}
\includegraphics[width=\hsize]{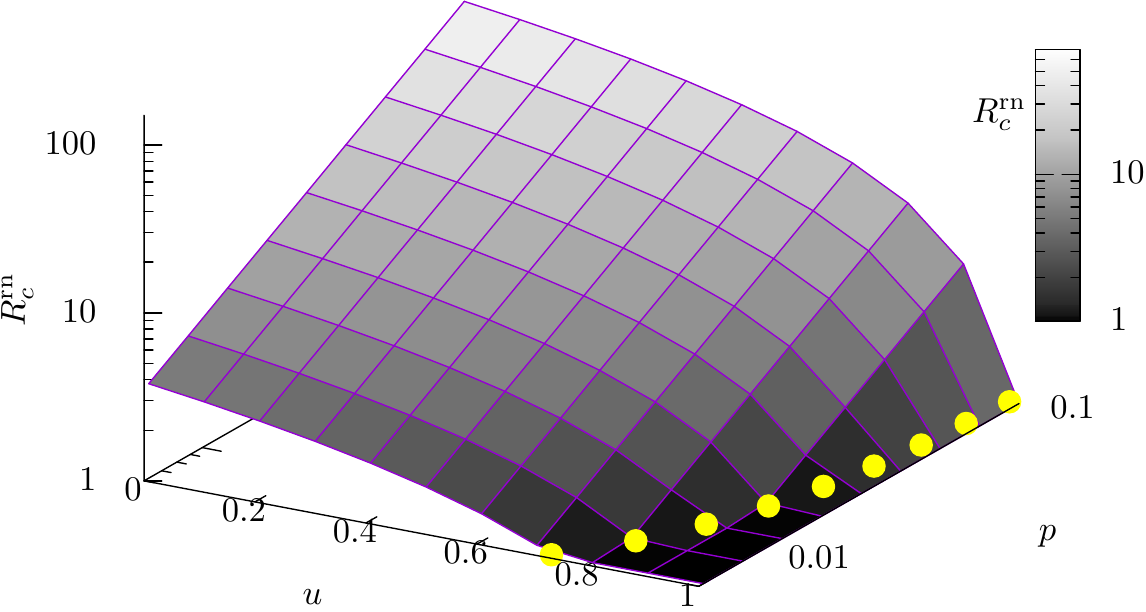}
\caption{The theoretically derived global reproduction number $R_c^{\rm rn}$ (Eq.~(\ref{eq:R_random})) as a function of the intervention rate $u$ and the mobility rate $p$ for a scale-free network with a degree distribution $p(k) \sim k^{-\gamma}$ with $\gamma = 2.1$, under the random intervention. The level of the global reproduction number $R_c^{\rm rn}$ is also indicated by gray-scale color. The value of $R_{c}^{\rm rn}$ increases monotonically with decreasing $u$ and increasing $p$.
 The yellow filled circles represent the critical intervention threshold $u_{c}^{\rm rn}$ as a function of $p$, which is theoretically derived from Eq.~(\ref{eq:uc_random}). The value of $u_{c}^{\rm rn}$ increases with $p$.}
\label{fig:R_theoretical_random}
\end{center}
\end{figure}

The global reproduction number $R_{c}^{\rm rn}$  for random intervention in a scale-free patch network is computed from Eq.~(\ref{eq:R_random}) and plotted as a function of the intervention rate $u$ and the mobility rate $p$ in Fig.~\ref{fig:R_theoretical_random}. The yellow filled circles represent the critical intervention threshold $u^{\rm rn}_{c}$ given by Eq.~(\ref{eq:uc_random}). As seen from Fig.~\ref{fig:R_theoretical_random}, $R_{c}^{\rm rn}$ decreases monotonically with the intervention rate $u$ and increases monotonically with the mobility rate $p$. The fact that the value of $u^{\rm rn}_{c}$ increases with $p$ suggests that more interventions for epidemic control are required when spatial movements of individuals are more active.

\begin{figure}[t]
\begin{center}
\includegraphics[width=\hsize]{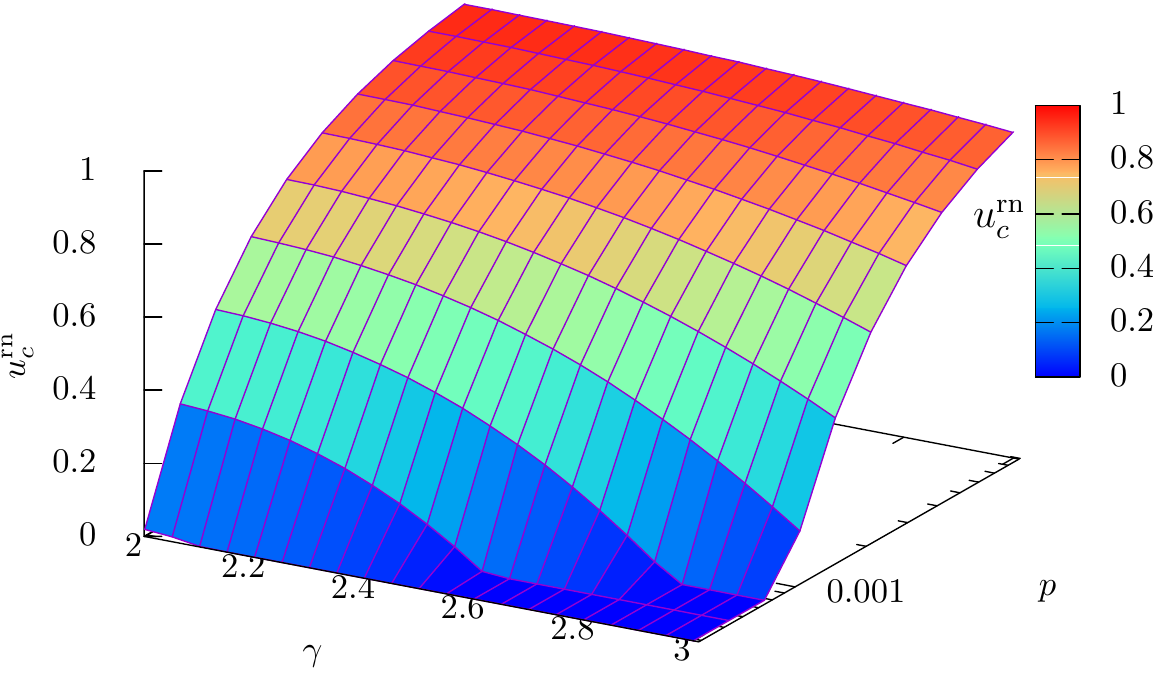}
\caption{The theoretically derived intervention threshold $u_c^{\rm rn}$ as a function of the degree exponent $\gamma$ and the mobility rate $p$. The level of the intervention threshold $u_c^{\rm rn}$ is also indicated by color. The intervention threshold increases with decreasing $\gamma$ and increasing $p$.}
\label{fig:uc_theoretical_random}
\end{center}
\end{figure}

The theoretical result in Eq.~(\ref{eq:uc_random}) also reveals the influence of the underlying mobility network on the intervention threshold. The connectivity of the scale-free patch network is varied with the degree exponent $\gamma$, which is typically in the range $2 < \gamma < 3$ \cite{barabasi2016network}. For a scale-free network with a degree distribution $p(k) \sim k^{-\gamma}$, the maximum degree $k_{\rm max}$ and the minimum one $k_{\rm min}$ satisfy the following relationship \cite{barabasi2016network}: 
\begin{equation}
k_{\rm max} =k_{\rm min} V^{\frac{1}{\gamma-1}}. \label{eq:kmax}
\end{equation}
Therefore, a smaller value of $\gamma$ means a larger difference between the maximum and minimum degrees. The intervention threshold given by Eq.~(\ref{eq:uc_random}) is plotted as a function of the degree exponent $\gamma$ and the mobility rate $p$ in Fig.~\ref{fig:uc_theoretical_random}. We can see that, for a fixed value of the mobility rate $p$, the intervention threshold $u_{c}$ increases with a decrease in $\gamma$. This result shows that more interventions are required for preventing a global outbreak in scale-free patch networks having a hub patch with a larger maximum degree. At $\gamma \simeq 2$, we have $k_{\rm max} \simeq V$ from Eq.~(\ref{eq:kmax}) and there exists a hub patch that connects to almost all other patches. Such a big hub patch easily causes a global outbreak and leads to a large critical intervention rate.

\subsubsection{Threshold for targeted intervention}
Next, we consider {\it targeted} intervention, which means that important patches are preferentially selected to be low-risk. Here we measure the importance of a patch using the degree centrality~\cite{freeman1978centrality}; the more connections a patch has, the more likely it is chosen as a low-risk patch. In this case, $q(k)$ should be a monotonically increasing function of $k$.

\begin{figure}[b]
\begin{center}
\includegraphics[width=0.9\hsize]{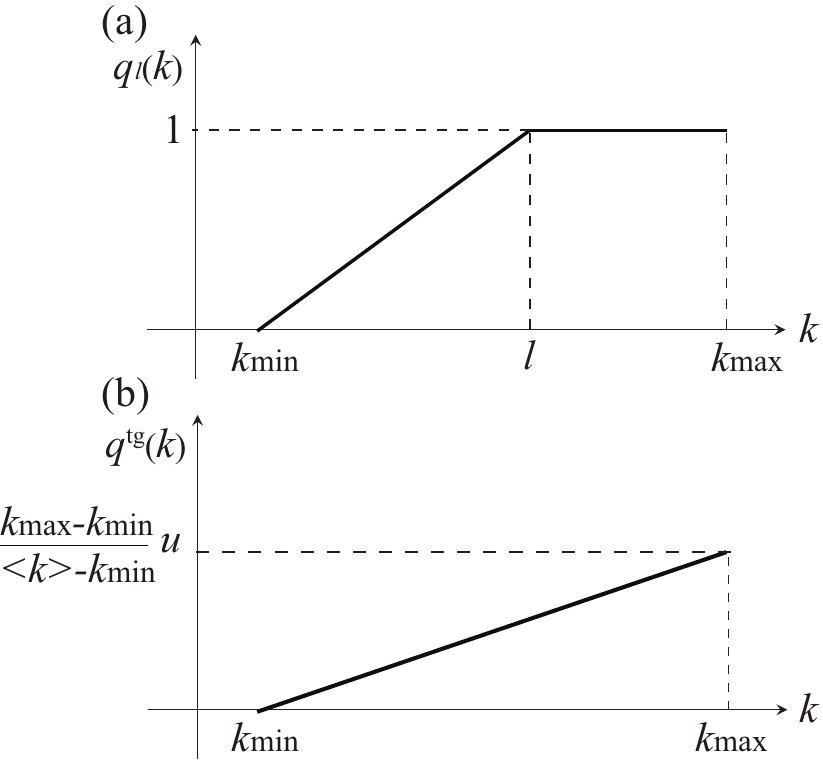}
\caption{(a) The piecewise function $q_l(k)$ defined in Eq.~(\ref{eq:q_l}). (b)  The probability density function $q^{\rm tg}(k)$ for $u<\hat{q}_{k_{\rm max}}$ in Eq.~(\ref{eq:q_targeted}).}
\label{fig:Qkshape}
\end{center}
\end{figure}

As a candidate of such a function, we define a piecewise function $q_l(k)$ as shown in Fig.~\ref{fig:Qkshape}(a), represented as follows:
\begin{eqnarray}
q_{l}(k) &:=& 
\begin{cases}
\dfrac{k-k_{\rm min}}{l - k_{\rm min}} & \mbox{for}~k_{\rm min} \le k < l, \\
1 & \mbox{for}~l \le k \le k_{\rm max},
\end{cases} \label{eq:q_l}
\end{eqnarray}
where $l$ is a real value ranging between $k_{\rm min}$ and $k_{\rm max}$.
We define the expectation value of $q_l(k)$ with respect to $k$ as follows: 
\begin{eqnarray}
\hat{q}_l &:=& \int_{k_{\rm min}}^{k_{\rm max}} q_{l}(k)p(k)dk. \label{eq:A_l} 
\end{eqnarray}
For scale-free networks with $p(k)\sim k^{-\gamma}$, we can show that $\hat{q}_l$ is monotonically decreasing with increasing $l$. In the limit of $l\rightarrow k_{\rm min}$, $\hat{q}_l$ approaches the maximum value 1. When $l=k_{\rm max}$, $\hat{q}_l$ takes the minimum value $\hat{q}_{k_{\rm max}}$.

We define the probability density function $q(k)$ separately for the two cases of $u \ge \hat{q}_{k_{\rm max}}$ and $u < \hat{q}_{k_{\rm max}}$. If $u \ge \hat{q}_{k_{\rm max}}$, we can find $k^* \in [k_{\rm min},k_{\rm max}]$ such that $\hat{q}_{k^*}=u$, satisfying Eq.~(\ref{eq:q_condition}). Therefore, we use $q_{k^*}$ as $q(k)$. Otherwise, the piecewise function with any $l$ does not satisfy Eq.~(\ref{eq:q_condition}). In this case, we use a non-piecewise function as shown in Fig.~\ref{fig:Qkshape}(b). The probability density function $q(k)$ for the targeted intervention is defined as follows:
\begin{eqnarray}
q^{\rm tg}(k) = 
\begin{cases}
q_{k^*}(k) \quad (u \ge \hat{q}_{k_{\rm max}}),\\
 \frac{k - k_{\rm min}}{\ave{k} - k_{\rm min}} u  \quad (u < \hat{q}_{k_{\rm max}}).
\end{cases} \label{eq:q_targeted}
\end{eqnarray}
We can show that the latter case also satisfies the requirements for the probability density function, Eqs.~(\ref{eq:q_range})-(\ref{eq:q_condition}), as follows:
\begin{eqnarray}
q^{\rm tg}(k) &=& \frac{k - k_{\rm min}}{\ave{k} - k_{\rm min}} u \nonumber \\
&\le & \frac{k_{\rm max} - k_{\rm min}}{\ave{k} - k_{\rm min}} \hat{q}_{k_{\rm max}} \nonumber \\
&=& \frac{k_{\rm max} - k_{\rm min}}{\ave{k} - k_{\rm min}} \int_{k_{\rm min}}^{k_{\rm max}} \dfrac{k-k_{\rm min}}{k_{\rm max} - k_{\rm min}} p(k)dk \nonumber \\
&=& 1, \nonumber
\end{eqnarray}
\begin{eqnarray}
\int_{k_{\rm min}}^{k_{\rm max}} q^{\rm tg}(k)p(k)dk &=& \int_{k_{\rm min}}^{k_{\rm max}} \frac{k - k_{\rm min}}{\ave{k} - k_{\rm min}} up(k)dk=u. \nonumber
\end{eqnarray}

\begin{figure}[t]
\begin{center}
\includegraphics[width=\hsize]{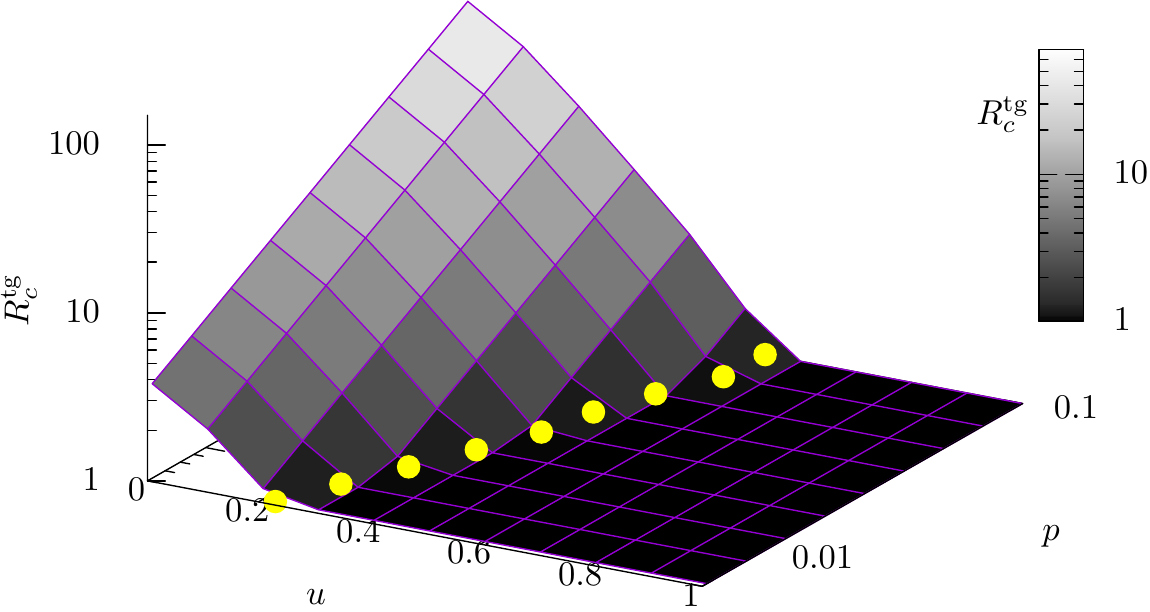}
\caption{The theoretically derived global reproduction number $R_c^{\rm tg}$ as a function of the intervention rate $u$ and the mobility rate $p$ for a scale-free network with a degree distribution $p(k) \sim k^{-\gamma}$ with $\gamma = 2.1$, under the targeted intervention. 
The level of the global reproduction number $R_c^{\rm tg}$ is also indicated by gray-scale color. $R_{c}^{\rm tg}$ increases with decreasing $u$ and increasing $p$.
 The yellow filled circles represent the critical intervention threshold $u_{c}^{\rm tg}$ as a function of $p$, which is theoretically derived from Eqs.~(\ref{eq:R_mix2}) and (\ref{eq:q_targeted}). The value of $u_{c}^{\rm tg}$ increases with $p$.}
\label{fig:R_theoretical_targeted}
\end{center}
\end{figure}

For the probability density function $q(k)=q^{\rm tg}(k)$, the global reproduction number $R^{\rm tg}_c$ is obtained as Eq.~(\ref{eq:R_mix2}) which depends on $\phi_2$ in Eq.~(\ref{eq:phi2}) where $\lbrack k^2 \rbrack$ and $\lbrack k \rbrack$ are computed from Eq.~(\ref{eq:bracket_k_cont}) with Eq.~(\ref{eq:q_targeted}). The global reproduction number $R_{c}^{\rm tg}$ for a scale-free patch network is shown as a function of the intervention rate $u$ and the mobility rate $p$ in Fig.~\ref{fig:R_theoretical_targeted}. As in the random intervention case, $R_{c}^{\rm tg}$ increases with decreasing $u$ and increasing $p$. We can obtain the critical intervention threshold $u^{\rm tg}_c$ by numerically solving $R_c^{\rm tg}=1$ with respect to $u$. The yellow filled circles represent the values of $u^{\rm tg}_{c}$ which increase with $p$. 
By comparing Fig.~\ref{fig:R_theoretical_targeted} with Fig.~\ref{fig:R_theoretical_random}, it can be visually confirmed that the critical intervention threshold for targeted intervention ($u^{\rm tg}_c$) is much smaller than that for random intervention ($u^{\rm rn}_c$).

\begin{figure}[t]
\begin{center}
\includegraphics[width=\hsize]{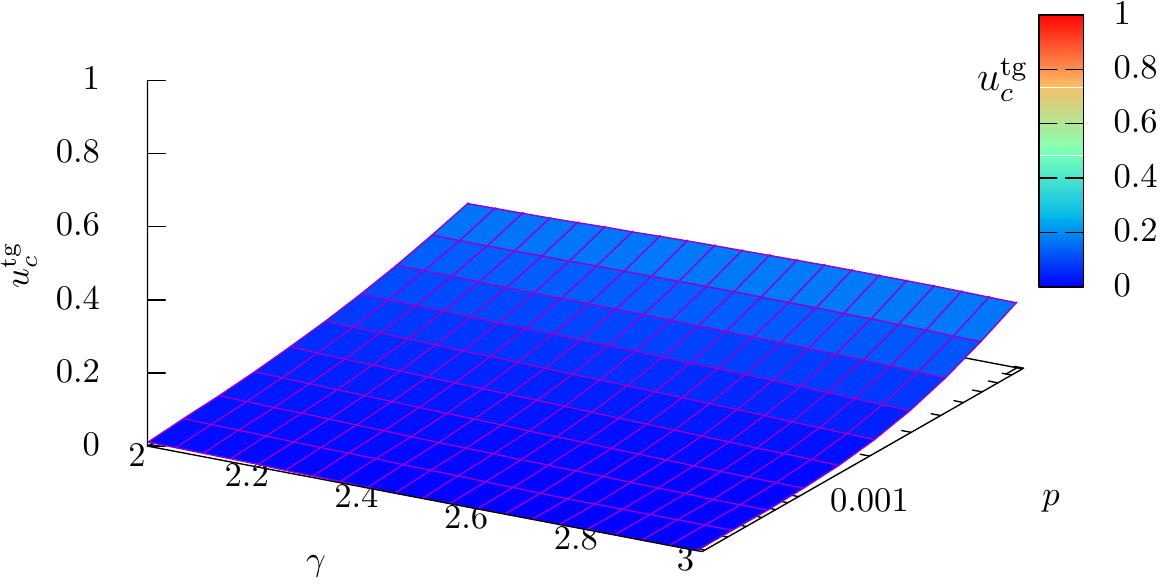}
\caption{The theoretically derived intervention threshold $u_c^{\rm tg}$ as a function of the degree exponent $\gamma$ and the mobility rate $p$. The level of the intervention threshold $u_c^{\rm tg}$ is also indicated by color.}
\label{fig:uc_theoretical_targeted}
\end{center}
\end{figure}

The critical intervention threshold $u^{\rm tg}_c$ is obtained by solving $R_c=1$ with respect to $u$ using Eq.~(\ref{eq:R_mix2}) and Eq.~(\ref{eq:q_targeted}). The threshold is plotted as a function of the degree exponent $\gamma$ and the mobility rate $p$ in Fig.~\ref{fig:uc_theoretical_targeted}. In contrast to the random intervention case, there is no remarkable change in $u_c$ with increasing $\gamma$ for a fixed value of $p$. This result implies that a very small intervention rate with which a few hub patches are changed to low-risk ones is sufficient for preventing global outbreaks.



\subsubsection{Comparison of the intervention thresholds}
The global reproduction number $R_c$ in Eq.~(\ref{eq:R_mix2}) is different between the random and targeted interventions, because $\phi_2$ depends on $q(k)$. A smaller value of $R_c$ for the same intervention rate $u$ means a more effective intervention strategy. We show that the targeted intervention is more effective than the random one. It is sufficient to prove the following inequality: 
\begin{eqnarray}
\Delta\phi_{2}(u) &:=& \phi_{2}^{\rm tg} - \phi_{2}^{\rm rn} \nonumber \\
 &=& \frac{1}{\langle k \rangle^2}\int_k (k^2-k)(q^{\rm tg}(k)-q^{\rm rn}(k))p(k)dk \nonumber \\ 
& \ge & 0, \label{eq:difference}
 \label{eq:dphi}
\end{eqnarray}
where $\phi_{2}^{\rm rn}$ and $\phi_{2}^{\rm tg}$ denote $\phi_2$ (Eq.~(\ref{eq:phi2})) for $q(k)=q^{\rm rn}(k)$ and $q(k)=q^{\rm tg}(k)$, respectively. We first deal with the case of $u\ge \hat{q}_{k_{\rm max}}$ and then that of $u < \hat{q}_{k_{\rm max}}$.

\begin{figure*}[t]
\begin{center}
\includegraphics[width=\hsize]{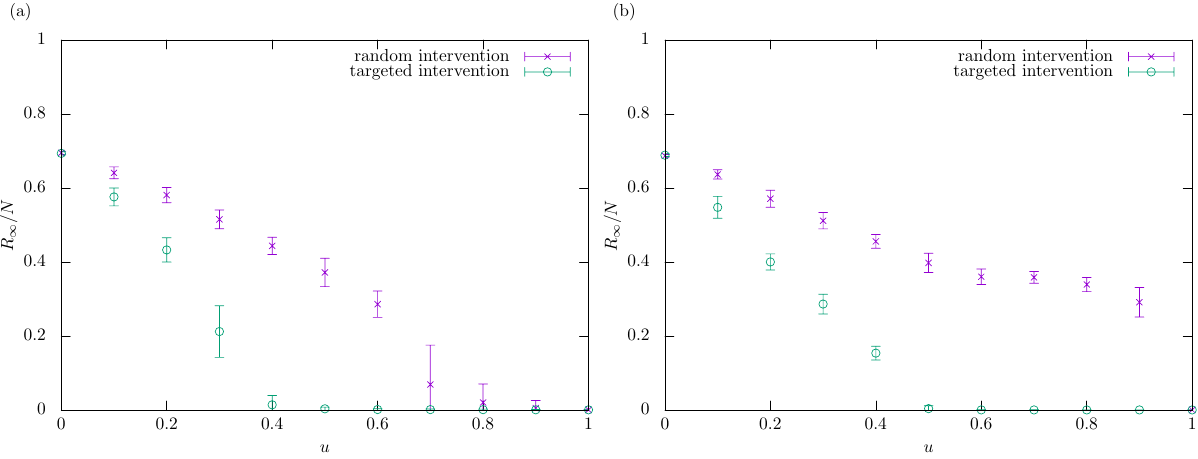}
\caption{Numerical results for the final epidemic size $R_{\infty}/N$ with different values of the intervention rate $u$. The mobility rate is fixed at $p=0.05$. The crosses and open circles indicate the average values over 50 simulations for the random and targeted interventions, respectively. The error bar indicates the standard deviation. (a) Scale-free patch networks generated with the configuration model \cite{catanzaro2005generation}. (b) The US airport network \cite{colizza2007reaction}.}
\label{fig:gar-u}
\end{center}
\end{figure*}


First, we assume $u \ge \hat{q}_{k_{\rm max}}$. We can evaluate $\Delta\phi_{2}(u)$ as follows: 
\begin{eqnarray*}
\ave{k}^{2}\Delta\phi_{2}(u) 
&=& \int_k (k^{2} - k) (q^{\rm tg}(k)-u)p(k)dk \\
&=& \int_k (k^{2} - k) (q_{k^*}(k) - \hat{q}_{k^*})p(k)dk \\
&=& \int_{k} \left( (k^{2} - k) - \ave{k^{2}-k} \right) q_{k^*}(k)p(k)dk.
\end{eqnarray*}
From Eq.~(\ref{eq:q_l}), the last term is equivalent to
\begin{eqnarray*}
&& \int_{k_{\rm min}}^{k^*} \left( (k^{2} - k) - \ave{k^{2}-k} \right)\frac{k- k_{\rm min}}{k^*- k_{\rm min}}p(k)dk \\
&& + \int_{k^*}^{k_{\rm max}} \left( (k^{2}-k) - \ave{k^{2}-k}\right)p(k)dk \\
&& =\int_{k_{\rm min}}^{k^*} \left( (k^{2} - k) - \ave{k^{2}-k} \right)\frac{k- k_{\rm min}}{k^*- k_{\rm min}}p(k)dk \\
&& - \int_{k_{\rm min}}^{k^*} \left( (k^{2}-k) - \ave{k^{2}-k}\right)p(k)dk \\
&& =\int_{k_{\rm min}}^{k^*}\left( (k^{2}-k) - \ave{k^{2}-k}\right) \frac{k-k^*}{k^*-k_{\rm min}}p(k)dk. \\
\end{eqnarray*}

\begin{figure*}[t]
\begin{center}
\includegraphics[width=\hsize]{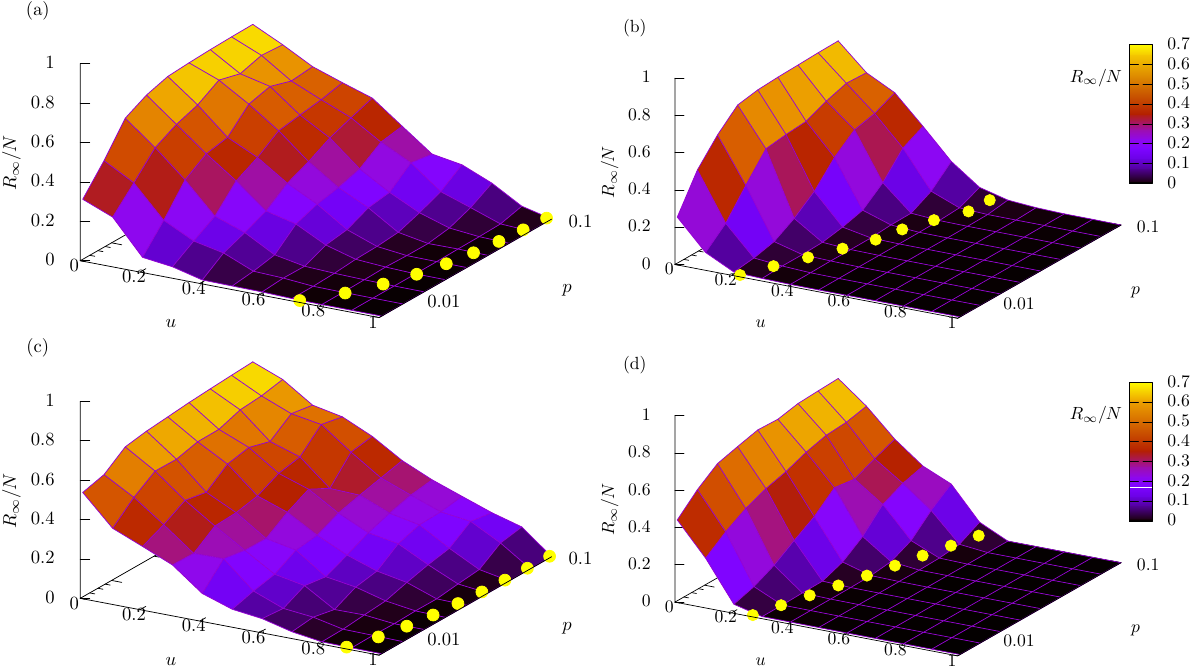}
\caption{Numerical results for the final epidemic size $R_{\infty}/N$ plotted against the intervention rate $u$ and the mobility rate $p$. The average value over 50 simulations is plotted for each parameter condition. The level of the epidemic size $R_{\infty}/N$ is also indicated by color. The yellow filled circles indicate theoretically obtained critical intervention thresholds $u_c$ for different values of $p$. (a) Random intervention in synthetic scale-free patch networks. (b) The same as (a), but for targeted intervention. (c) Random intervention in the US airport network. (d) The same as (c), but for targeted intervention. The theoretical values of $u_c$ are computed from Eq.~(\ref{eq:uc_random}) for (a) and (c), and from Eqs.~(\ref{eq:R_mix2}) and (\ref{eq:q_targeted}) for (b) and (d).
}
\label{ugareng}
\end{center}
\end{figure*}

Now we define the following functions:
\begin{eqnarray}
b(k) &:=& \left( (k^{2}-k) - \ave{k^{2}-k}\right)p(k),\\
B_l &:=& \int_{k_{\rm min}}^l b(k)(k-l)dk, \label{eq:B_l}
\end{eqnarray}
where $b(k)$ is a monotonically increasing function of $k$ for a scale-free network, satisfying $b(k_{\rm min})<0$, $b(k_{\rm max})>0$, and $\int_k b(k)dk=0$. Using these functions, we can represent $\Delta\phi_{2}(u)$ as follows:
\begin{eqnarray}
\Delta\phi_{2}(u) &=& \frac{B_{k^*}}{\ave{k}^{2}(k^*-k_{\rm min})}. \label{eq:difference2}
\end{eqnarray}
Therefore, the inequality (\ref{eq:difference}) holds if $B_{k^*}$ is non-negative. From Eq.~(\ref{eq:B_l}), we have
\begin{eqnarray}
\frac{dB_l}{dl} &=& -\int_{k_{\rm min}}^l b(k)dk, \label{eq:dBdl}\\
\frac{d}{dl}\left(\frac{dB_l}{dl}\right) &=& -b(l). \label{eq:dBdldl}  
\end{eqnarray}
From the monotonicity of $b(k)$, $dB_l/dl$ in Eq.~(\ref{eq:dBdl}) is a unimodal function. Therefore, we have 
\begin{eqnarray}
\frac{dB_l}{dl} &\ge & \min \left\{ -\int_{k_{\rm min}}^{k_{\rm min}} b(k), -\int_{k_{\rm min}}^{k_{\rm max}} b(k) \right\}=0.
\end{eqnarray}
Hence, $B_l$ is a monotonically increasing function of $l$. From $B_{k_{\rm min}}=0$, we obtain $B_{k^*} \ge 0$. From Eq.~(\ref{eq:difference2}), $\Delta\phi_{2}(u) \ge 0$ is satisfied.


Next, we assume $u < \hat{q}_{k_{\rm max}}$. From Eq.~(\ref{eq:difference}), it follows
\begin{eqnarray}
& & \ave{k}^{2} \Delta\phi_{2}(u) \nonumber \\
&=& \int_{k} (k^{2} - k) (q^{\rm tg}(k)-u)p(k)dk \nonumber \\
&=& \int_k (k^2-k) \left(\frac{k-k_{\rm min}}{\ave{k}- k_{\rm min}} u-u\right)p(k)dk\nonumber \\
&=& \frac{u \left( \ave{k^{3} - k_{\rm min }k^{2}}  - \ave{k^{2} - k_{\rm min }k} \right) }{\ave{k}- k_{\rm min}} - u\ave{k^{2} - \ave{k}} \nonumber \\
 &=& \frac{u}{\ave{k} - k_{\rm min}}(\ave{k^{3}} - \ave{k^{2}} - \ave{k}\ave{k^{2}} + \ave{k}^{2}). \label{eq:3-dphi2-2}
\end{eqnarray}
From $k \ge k_{\rm min} > 0$, we have $k(k-\ave{k})^{2} \ge 0$, yielding $\ave{k(k-\ave{k})^{2}} \ge 0$. This yields $\ave{k^3} \ge 2\langle k \rangle \langle k^2 \rangle -\langle k \rangle^3 \ge 0$. From this property, we can evaluate Eq.~(\ref{eq:3-dphi2-2}) as follows:
\begin{eqnarray}
&& \ave{k}^{2} \Delta\phi_{2}(u) \nonumber \\
&\ge& \frac{u}{\ave{k}-k_{\rm min}} (\ave{k}\ave{k^{2}} - \ave{k}^3 - \ave{k^{2}} + \ave{k}^{2}) \nonumber \\
&=& \frac{u}{\ave{k}-k_{\rm min}} (\ave{k}-1)(\ave{k^{2}} - \ave{k}^{2}) \nonumber \\
&=& \frac{u}{\ave{k}-k_{\rm min}} (\ave{k}-1) \ave{(k-\ave{k})^{2}} \nonumber \\
&\ge& 0.
\end{eqnarray}

Therefore, $\Delta\phi_{2}(u) \ge 0$ holds. From Eq.~(\ref{eq:3-dphi2-2}), we find that $\Delta\phi_{2}(u)$ is a monotonically increasing function of $u$.



\subsection{Numerical validation \label{sebsec:numerical}}
We numerically study the effect of the local interventions on the final epidemic size. The final epidemic size is measured by the ratio of individuals who have experienced the disease during an outbreak period, given by $R_{\infty} / N$ where $R_{\infty}$ equals to $\sum_j R_j$ after the outbreak. 
Due to the finiteness of the number of degrees in simulations, we used a discretized version of Eq.~(\ref{eq:q_targeted}) for the targeted intervention.

The average values of the final epidemic size over 50 simulations are plotted against the intervention rate $u$ in Fig.~\ref{fig:gar-u}(a) for synthetic scale-free patch networks generated with the configuration model \cite{catanzaro2005generation} and in Fig.~\ref{fig:gar-u}(b) for the U.S. airport network representing the connectivity of flight routes between 500 major airports in United States \cite{colizza2007reaction}. We see that, in both networks, the targeted intervention is much more effective than the random intervention as it requires a much lower intervention rate for containment of epidemics.

In Fig.~\ref{ugareng}, the numerical results of the final epidemic size are shown for variation of the intervention rate $u$ and the mobility rate $p$. Figures~\ref{ugareng}(a) and (b) correspond to the results for random and targeted interventions in synthetic scale-free patch networks, respectively. A comparison between these two figures obviously shows that the targeted intervention is more effective than the random intervention for reducing the epidemic size. The same property is confirmed in Figs.~\ref{ugareng}(c) and (d), which correspond to random and targeted interventions in the US airport network, respectively. In all the cases, the final epidemic size decreases with increasing $u$ and increases with increasing $p$. The theoretical values of the intervention threshold $u_c$ are superimposed as yellow filled circles, indicating that they are in good agreement with the thresholds which are recognized from the numerical results.

\section{Conclusion and Discussion \label{sec:conclusion}}
In the present study, we have analyzed the intervention threshold in SIR metapopulation models with scale-free patch connectivity, consisting of high-risk and low-risk patches. Under the assumption that a high-risk patch is changed to a low-risk patch by reducing the local reproduction number by an intervention, we have compared the effectiveness of random and targeted interventions through theoretical and numerical analyses. The theoretical results have shown that the intervention targeted to high-degree patches is more effective than the random intervention. They have been validated by the numerical simulations using the synthetic scale-free networks and the realistic US airport network. 
Our result indicating the effectiveness of the targeted intervention for SIR metapopulation models is consistent with that for SIS metapopulation models in a similar framework \cite{tanaka2014random}. As the global reproduction number is expressed as a function of the intervention rate and the mobility rate, one can calculate the critical intervention threshold for a given mobility rate and estimate the minimum effort for containment of epidemics. 
We have found that a higher human mobility rate leads to a larger intervention threshold. This finding suggests that travel restrictions are effective, especially when using targeted intervention.
We have also revealed that the intervention threshold is larger for a scale-free patch network with a smaller degree exponent. This result implies that an existence of a very big hub patch increases the difficulty of preventing global outbreaks.

The framework for examining intervention strategies in this study has a potential to be extended to more realistic cases in terms of human mobility patterns and intervention strategies. There are other types of human mobility patterns, such as recurrent (commuting) mobility \cite{belik2011recurrent,balcan2011phase, panigutti2017assessing,gomez2018critical, granell2018epidemic} and adaptive mobility \cite{meloni2011modeling,wang2012safety}. Moreover, human mobility networks can be better estimated from higher-resolution data such as real-world traffic network data \cite{merler2009role,balcan2010modeling}, mobile phone data \cite{wesolowski2012quantifying,tizzoni2014use,panigutti2017assessing}, and GPS data \cite{vazquez2013using}. It is also intriguing to test other intervention strategies, such as those based on other network centralities, because the important patches are not necessarily high-degree ones. Another strategy is to combine the intervention to local patches and travel restrictions. It would be possible that the optimal intervention strategy is different depending on how to evaluate the epidemic outcome. Therefore, an appropriate assessment of the social impact of global epidemics from microscopic and macroscopic levels is becoming increasingly important \cite{ball2015seven,massaro2018resilience}.

\appendix

\end{document}